# Extreme Value Statistics and Error Correcting Codes

**S. Rouhani & J. Davoudi**
*Department of Physics, Sharif Univ. of Technology &*
*Institute for Studies in Theoretical Physics & Mathematics (I.P.M.)*
*P. O. Box 19395-5746*
*Tehran, Iran*

## Abstract

Derrida's model can be used for construction and transmission of error correcting codes, in an optimal way. We use the extreme statistics method to derive the optimal signal to noise ratio.

## Introduction

A given message can be coded into a sequence of N bits $\vec{\sigma} = \{\sigma_1, \cdots, \sigma_N\}, \sigma_i = \pm 1$. Once transmitted through a communication channel, we may receive a different sequence say $\vec{\sigma}^1$, depending on the statistical properties of the channel. Clearly decoding may result in obtaining a corrupted message in place of the original message. However if we had allowed for some redundancy in the original coding, allowing for the errors resulting from the communication channel one could still recover the original message, uncorrupted. More precisely, instead of the original sequence we transmit:

$$J_k^{in} = C_{i_{k_1} \cdots i_{k_p}}^{(k)} \sigma_{i_{k_1}} \cdots \sigma_{i_{k_p}}, \quad k = 1, \cdots, M \tag{1}$$

We have thus replaced the original sequence with a longer sequence $(M > N)$, the price we have paid for immunity from information corruption, is that we transmit a longer sequence. The connectivity matrix $C_{i_{k_1} \cdots i_{k_p}}^{(k)}$ which has elements zero, or one, is in fact the "rule" for constructing the error correcting code. The ratio $R = \dfrac{N}{M}$, called the rate of the code, counts its redundancy. The tasks which face us are, first determine the connectivity matrix in such a way that $R$ is maximised, second decode the transmitted message in an efficient manner in order to recover the original message. Shannon's theorem sets a limit on the rate, in terms of the channel capacity. The channel capacity is the maximum information per unit time which can be transmitted through the channel [1].

In general, when a signal $\sigma$, is sent through a channel, the output will be a real number $X$. Let $Q(u|\sigma)\,du$ be the probability of observing the output between $u$ and $u + du$ given an input $\sigma$. A channel is called memoryless if the noise is independent of the bit transmitted:

$$Q(\vec{x}|\vec{\sigma}) = \prod_i Q(x_i|\sigma_i) \tag{2}$$

Shannon shows that for a memoryless channel and gaussion noise the channel capacity is given by:

$$C = \frac{1}{2\ell n2} \ell n\left(1 + \frac{V^2}{\omega^2}\right) \tag{3}$$

Where $\frac{V}{\omega}$ is the signal to noise ratio. According to Shannon the rate $R$ is always less than or equal to $C$. However Shannon shows that ideal codes which saturate the bound exist. Spin glass theory provides us with a way of constructing such codes.

Sourlas [2] has pointed out that the p-spin hamiltonian of Derrida [3] provides a good example of error free coding. The original message sequence $\overset{t}{\sigma}$ is the ground state of a p-spin hamiltonian. Thus the task of decoding is equivalent to finding the ground state of p-spin hamiltonian. The hamiltonian is constructed in such a way that it operates in the ferromagnetic phase, thus there exist a unique global ground state, in the absence of the channel noise. In the presence of channel noise the bonds take on random values, thus the existance of a global ground state becomes questionable. However, the parameters are tuned in such a way that the system remains in the ferromagnetic phase, therefore despite the noise there is still a chance of recovering the original message.

More precisely, the input is the sequence defined by equation (1), the output is a sequence of $M$, real numbers $J_k^{out}$, which are random and obey the distribution $Q\left(J_k^{out}/J_k^{in}\right)$ satisfying the constraint set out in equation (2).

Now the probability that any sequence $\sigma$ is the original message conditional on the output $J_k^{out}$, $P\left(\sigma|J^{out}\right)$ is given by [4];

$$\ell nP\left(\sigma|J^{out}\right) = const - H(\sigma) + \sum_{k=1}^{M} C_{i_{k_1}}^{(k)} \cdots {}_{i_{k_p}} B_k \sigma_{i_{k_1}} \cdots \sigma_{i_{k_p}} \tag{4}$$

where $H(\sigma)$ gives the bias in the original code, this may be constant. The coefficients $B_k$ are given by:

$$B_k = \frac{1}{2}\ln\frac{Q\left(J_k^{out}/1\right)}{Q\left(J_k^{out}/-1\right)} \tag{5}$$

Equation (4) is indeed the defining relation for Derrida's p-spin model. The most probable code is the ground state of the hamiltonian defined by (4). As a concrete case let us consider the fully connected version where all $C^{(k)}$ are set equal to 1. Let all codes be equally probable, $H(\sigma) = 0$, and let the channel be gaussian. Then, we have the following hamiltonian:

$$H = \sum_{\{i\}} J_{i_1\cdots i_p}\sigma_{i_1}\cdots\sigma_{i_p} \tag{6}$$

Where $J_{i_1\cdots i_p}$ are output values with a normal distribution:

$$P(J_{ss}) \sim \exp\left\{ -\frac{J_{ss} - J_s N^{P - \frac{1}{p!}}}{2\frac{N^{p-1}}{p!}\omega^2} \right\}$$

(7)

Where we have assumed that $1 \ll p \ll N$. Here the original code is assumed to be the trivial signal where all $\sigma_i = 1$. This is not a resticting assumption since, hamiltonian (6) is gauge invariant under the transformation $\sigma_i \rightarrow \sigma_i \varepsilon_i$, $\varepsilon_i = \pm 1$. Thus all codes are equivalent.

The procedure for decoding is equivalent to fiding the ground state of the hamiltonian (6). Of course the rate is very poor:

$$R = \frac{p!}{N^{p-1}}$$

(8)

As is well known spin glass hamiltonians such as Eq. 6 can have many minima, thus the task of decoding is formidable, unless the system operates in its ferromagnetic phase. In this case there should exist a unique ground state. The phase diagram of Derrida's model in preseme of a ferromagnetic interaction was derived in [5, 6]. Indicating that there does indeed exist a region in phase space, even for finite $N$, where a unique ferromagnetic phase exists. The critical value of signal to noise ratio is:

$$\frac{V}{\omega} > (2 \ln 2)^{\frac{1}{2}}$$

(9)

In this paper we aim to derive this result using extreme value statistics. A much more powerful tool which readity yields the result.

### Extreme Value Statistics

Consider $M$ independent identically distributed random variables $E_i, i = 1, L, M$, such that the probability distribution $P(E)$ decays faster than any power in infinity. What is the distribution of $E^*$ the minimum value of $E_i$, for large $M$?

This is clearly a statistics which is suited to the study of the lowest energy levels in disordered systems [7]. The cumulative probability $p_<(E)$ is defined as

$$P_<(E) = \int_{-\infty}^{E} dx \, p(x)$$

(10)

The distributions of $E^*$ is

$$P_M(E^N) = M \, P(E^*)(1 - p_<)^{M-1} = -\frac{d}{dE^*}\left[ p_>(E^*) \right]^M$$

(11)

For large $M$, the minimum $E^*$ will be negative and large. Thus $P_M(E^*)$ will have a peak near $E_c$; determined by the equation $M P_<(E_c) = 1$.

Now for our model hamiltonian $M = 2^N$ is indeed very large. Thus we can find the distribution of the ground state energy using the method above and the fact that for large $P$, Derrida's model is approximated by the random energy model [3]:

$$Z = \sum_{i=1}^{M} e^{-\beta E_i}, \; p(E_i) \sim e^{\frac{E_i^2}{2N}} \tag{12}$$

This immediately leads to the value:

$$E_c = -N(2\ell n2)^{\frac{1}{2}} \tag{13}$$

Now if a ferromagnetic term is present the distribution of the energies will be slightly different. The model is still defined by (11) except for one energy level with a different distribution due to the presence of $J_o$:

$$p(E_0) \sim e^{-\frac{(E+NJ_0)^2}{2N\omega^2}} \tag{14}$$

We thus need to modify equation (10) to take account of the fact that one energy level behaves differently:

$$p(E^*) = -\frac{d}{dE^*} \left[ P_<(E^*)^{M-1} P_>^{J_0}(E^*) \right] \tag{15}$$

Where $P_>^{J_0}$ is defined using the normal distribution with a mean at $-NJ_o$ rather than zero. We now find that the distribution $P(E^*)$ still has one peak at $E_c$ provided that

$$\frac{J_o}{w} < (2\ell n2)^{\frac{1}{2}} \tag{16}$$

Where as beyond this value, the distribution will have two peaks, with the lower peak at $-NJ_o$, that is the ferromagnetic phase. Therefore for efficient decoding one must have $\frac{J_o}{w} \geq (2\ell n2)^{\frac{1}{2}}$.

Let us next consider Derrida's model, with a binary distribution for the $J_{i_1 L i_p}$, namely

$$p(J_{o_i}) = \frac{1+\alpha m}{2} \delta(j_{o_i} - \alpha j_o) + \left( \frac{1-\alpha m}{2} \right) \delta(j_{o_i} + \alpha J_o)$$

where $\alpha = \sqrt{N^{1-p} p!}$. This is more appropriate to a digital channel, again we consider the fully connected case. We can show that the distribution for energy factorizes in the limit of large $p$, and zero magnetization:

$$f(E_1 \cdots E_M) = \prod_\alpha f(E_\alpha)$$

where

$$f(E) = \int \frac{d\lambda}{2\pi} e^{i\lambda E} \times (Cos\alpha\lambda)^N$$

where $x = \begin{pmatrix} N \\ p \end{pmatrix}$. Expanding we have alternatively:

$$f(E) = \left(\frac{1}{2}\right)^x \sum_{r=0}^{x} \begin{pmatrix} x \\ r \end{pmatrix} \delta(E + \alpha J(2r - x))$$

Which is approximated by a normal distribution of zero mean and variance $J^2 N$. We can thus repeat last section's argument to get the critical value

$$E_c = -N(2\ell n2)^{\frac{1}{2}}.$$

and

$$-mJ_c = (2\log 2)^{\frac{1}{2}}.$$

## Concluding Remarks

We have observed that extreme statistics offers a very powerful tool for dealing with systems with random energy levels. In particular extreme statistics immediately gives the threshold values for operation of Derrida's model as an error correcting code. The calculation equally well works in gaussian and digital channels.

Fully connected Derrida's model can saturate Shannon's limit, but this is stictty true only in the limit of large $N$, or vanishing capacity which is not practical.